\documentclass[prl,aps,twocolumn,superscriptaddress,showpacs,amsmath,amssymb,showkeys]{revtex4}

\usepackage{graphicx}
\usepackage{graphicx}
\begin{document}
\title{Magnetization Process of Single Molecule Magnets at Low 
Temperatures}
\author{Julio F. Fern\'andez}
\affiliation{ICMA, CSIC and Universidad de Zaragoza, 50009-Zaragoza, Spain}
\email[E-mail address: ] {JFF@Pipe.Unizar.Es}
\author{Juan J. Alonso}
\affiliation{F\'{\i}sica Aplicada I, Universidad de M\'alaga,
29071-M\'alaga, Spain}

\begin{abstract}     
We show that correlations established before quenching to very low 
temperatures, later drive the magnetization process of systems of
single molecule magnets, after a magnetic field is applied at $t=0$. 
We also show that in SC lattices,
$m\propto \sqrt{t}$, as observed in Fe$_{8}$, but only for $1+2\log 
_{10}(h_d/h_w)$ time decades, where
$h_d$ is a nearest neighbor dipolar magnetic field and a spin 
reversal can occur only if the field on it is within
($-h_w,h_w$). However, the $\sqrt{t}$ behavior is not universal. For 
BCC and FCC lattices, $m\propto t^p$, but $p\simeq 0.7$. The
value to which $m$ finally levels off is also given.
\end{abstract}
\maketitle
Magnetic clusters, such as Fe$_8$ and Mn$_{12}$, that make up the 
core of large organometallic molecules, behave at low temperatures
as large single spins. Accordingly, these molecules have come to be 
known as single molecule magnets (SMM's) \cite{SD}. In
crystals, magnetic anisotropy energies $U$ inhibit magnetic 
relaxation of SMM's, which can consequently take place at very small
temperatures only through magnetic quantum tunneling (MQT). Dipolar 
interactions play then an essential role. They can give rise,
upon tunneling, to Zeeman energy changes of nearly 1 K. This exceeds 
by many orders of magnitude the ground state tunnel splitting
energy
$\Delta$ that would follow from perturbations by higher anisotropies 
for Fe$_8$ and Mn$_{12}$ \cite{highan}.  Energy conservation
would make MQT, which has been observed experimentally
\cite{sangregorio}, impossible for the vast majority of spins in the 
system. Hyperfine interactions between the tunneling electronic
spins of interest and nuclear spins open up a fairly large tunneling 
window of energy
$\varepsilon_w$ such that tunneling can occur if the Zeeman energy 
change $2\varepsilon_h$ upon tunneling is not much larger than
$\varepsilon_w$ \cite{PS}. More precisely, the tunneling rate
$\Gamma^\prime$ for spins at very low temperature, is given by
\begin{equation}
\Gamma^\prime (\varepsilon_h) \simeq
\Gamma\;\eta (\varepsilon_h/\varepsilon_w),
\label{gamma}
\end{equation}
where $\Gamma$ is some rate (whose value is not important for our 
purposes), $\eta (x)\sim 1$ for $\mid x \mid < 1$, $\eta
(x)\sim 0$ for $x> 1$, and $\varepsilon_w\gg\Delta$. Other theories 
for MQT of SMM at very low temperatures have also been
proposed
\cite{chudno}. We adopt Eq. (\ref{gamma}) here, regardeless of theory 
or physical mechanism behind it. We let $\eta
(x)= 1$ for $\mid x\mid < 1$ and $\eta (x)=0$ for $x\ge 1$, and refer 
to $\varepsilon_w$ as the tunnel energy window.

The interesting early time relaxation $1-m\propto \sqrt {t}$ of an 
initially magnetized system has been predicted \cite{PS},
observed experimentally \cite{exp}, further explained \cite{vill}, 
and widely discussed \cite{comm}. An unpredicted related phenomenon
was later observed by Wernsdorfer et al. \cite{ww1}: the 
magnetization $m$ of a system of Fe$_8$ SMM's increases
as $\sqrt{t}$, where $t$ is the time after a weak magnetic field is 
applied to an initially unpolarized system. Conveniently, this
latter effect seemed to be independent of system shape. Interesting questions
arise: is this a universal effect to be found in all MQT experiments? 
If not, what does it depend on? How many time
decades does the $\sqrt{t}$ regime cover? What is the final steady 
state magnetization? No explanation or simulation that we know of
has been offered. We address these questions here.

We report Monte Carlo (MC) results that reproduce the $m\propto \sqrt 
{t}$ behavior of initially unpolarized systems. We
show that this arises from correlations that develop between spins 
and local magnetic dipolar fields, while cooling
to low temperatures, before finally quenching to experiment. 
Furthermore $m(t)$ depends on the cooling
protocol only through the final energy $-\varepsilon_a$ reached just 
before quenching.

The main results obtained follow. All energies and magnetic fields 
are given in terms of $\varepsilon_d$ and $h_d$, respectively,
where $-2\varepsilon_d$ is the energy of two S spins that lie on 
sites $a$ distance away, that point along the line joining their two
sites, $a$ is the side of a cubic unit cell,
$h_d=\varepsilon_d/(g\mu_BS)$, $g$ is the gyromagnetic ratio, and 
$\mu_B$ is the Bohr magneton. We also let
$\sigma$ stand for the rms value of $h$ for a disordered spin 
configuration \cite{Alonso2}.  After quenching
and applying a field $H\lesssim1$ at $t=0$,
\begin{equation}
m(t)\simeq  b\varepsilon_a \varepsilon_w H\sigma^{-3}F(\Gamma 
t,\sigma/\varepsilon_w,\sigma/h_0),
\label{mdet}
\end{equation}
where $b\simeq 4\sqrt{2/\pi}$, and
\begin{eqnarray}
F&\simeq & \Gamma t \;\;\text{ \hspace{1.2cm}  for } \Gamma t\lesssim 1\\
\label{Fdeta}
F&\simeq & 0.7(\Gamma t)^p \text{ \hspace{0.5cm}  for } 
1\lesssim\Gamma t\lesssim (\sigma/\epsilon_w)^{1/p}\\
F &\simeq &0.5\,\sigma\varepsilon_w^{-1} \text{ \hspace{0.5cm}  for }
(\sigma/\varepsilon_w)^{1/p}\lesssim
\Gamma t,
\label{Fdetb}
\end{eqnarray}
where, $h_0=2(2\pi )^2/3^{5/2}{\rho_v}$, $\rho_v$ is the number of 
spin sites per unit volume,
$p\simeq 0.5$, for simple cubic (SC) lattices, and $p\simeq 0.7$ for 
body centered cubic (BCC) and face centered cubic (FCC)
lattices. These results are inferred from MC simulations in which
the energy of the magnetic system is held constant in time as well as 
from arguments given below. For magnetic
systems that readily exchange energy with the lattice, for which the 
energy is not a constant of time, results are briefly discussed
in the closing remarks.
\begin{figure}
\includegraphics[width=8cm]{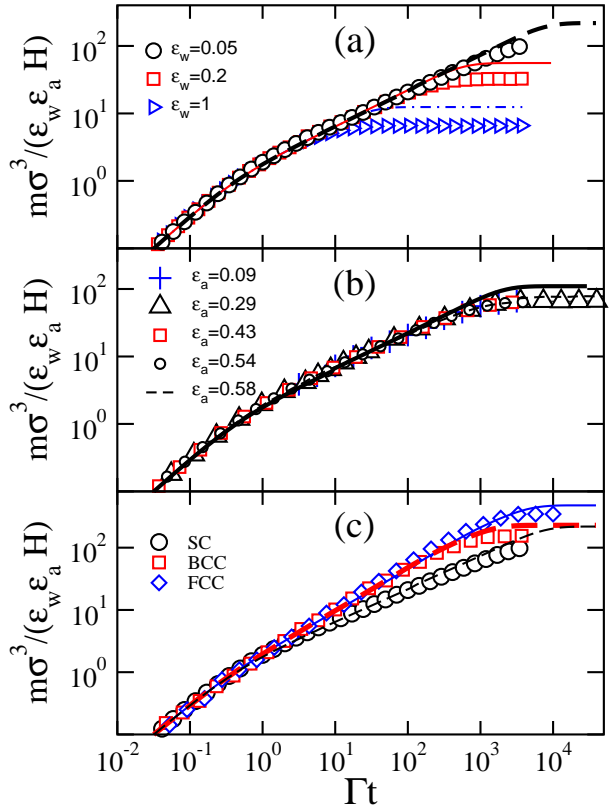}
\caption{(a) $m/\varepsilon_w$ versus $\Gamma t$ for an
applied external field $H=1$ for the shown values of $\varepsilon_w$. 
First, thermalization took place from an
initially disordered configuration up to the time when 
$\varepsilon_a$ reached the value $0.58$. The dashed, continuous, and
dot--dashed lines are for curves calculated from Eq. (\ref{mfinal}), 
for $\varepsilon_w=0.05$, $0.2$ and $1$, respectively. (b) Same
as in (a) but for the shown values of
$\varepsilon_a$ and
$\varepsilon_w=0.1$. The continuous line, for the solution from Eq. 
(\ref{mfinal}). (c) Same as in (a) and (b), but for
different lattices, with $\varepsilon_w=0.05$. For SC, BCC, and FCC 
lattices, previous partial thermalization took place at $T_a=10$,
$20$ and $60$, till $\varepsilon_a=0.31$, $0.28$, and $0.53$, 
respectively. The continuous, long dashed,
and short dashed lines are for curves calculated from Eq. 
(\ref{mfinal}) for FCC, BCC, and SC lattices, respectively. There are 
no
adjustable parameters.}
\label{fdet2}
\end{figure}
We first describe the simulations. We use the MC method to simulate 
magnetic relaxation of Ising systems of $\pm S$ spins, on simple
cubic lattices with periodic boundary conditions, that interact 
through magnetic dipolar fields and flip under rules to be specified
below \cite{Ising}. The system is first allowed to evolve towards 
thermal equilibrium at some ``high'' temperature $T_a$. We assume
$k_BT_a\gtrsim U/10$, which implies that spin reversals then take 
place mostly through classical thermal processes. Accordingly, spin
flips are then governed by detailed balance rules, and Eq. 
(\ref{gamma}) is not enforced. For reasons that are given below, we 
also
impose the restriction
$T_a\gtrsim T_0$, where $T_0$ is the long-range ordering temperature. 
One may think of this first process as a waiting stage that the
systems may have to undergo in the cooling process, before quenching 
to a lower temperature where a tunneling experiment (as in
Ref.~\onlinecite{ww1}) can be later performed. Let this first stage 
end at $t=0$ with sudden cooling of the system to a ``very
low'' temperature, that is, to a temperature below, roughly, 
$0.2U/(Sk_B)$ \cite{sangregorio,ree}. Accordingly, Eq. (\ref{gamma}) 
is
then enforced on all spin flips for $t>0$. As for detailed balance, 
we then proceed as follows. We assume that thermalization of a SMM
system with the lattice does not take place (i.e., the energy is 
constant) at very low temperatures (but see Ref. \cite{metes}). We
meet this condition by enforcing detailed balance but using an 
appropriately chosen pseudotemperature $T_u$. [From an expression
below Eq. (\ref{x2}),
$k_BT_u\approx \sigma^2 /2\varepsilon_a$. Note that $T_u\geq T_a$, 
since $-\varepsilon_a$ cannot be smaller than the
equilibrium energy at $T_a$]. We have checked that the mean energy is 
indeed constant under this rule. We do not report here
results we have obtained applying detailed balance rules with
$T<T_a$ (applicable to systems where thermal relaxation to the 
lattice takes place
\cite{metes}), but make a comment on them in the closing remarks.
MC results for the time evolution of $m/\varepsilon_w$ in SC lattice 
systems, after a field $H=1$ is applied upon quenching, are
shown in Fig. 1(a) for various values of $\varepsilon_w$. Before 
quenching, the system was thermalized at $T_a=10/k_B$ for some time
till the energy per spin reached the value $-0.58$. Clearly, $m$ 
scales with $\varepsilon_w$ up to a crossover time of, roughly,
$10\Gamma ^{-1}/\varepsilon_w^2$, where $m$ levels off. Within the 
time range $1\lesssim \Gamma t \lesssim
10/\varepsilon_w^2$, $m\propto \sqrt{t}$. Monte Carlo results that 
show how $m$ scales with $\varepsilon_a$ are exhibited in
Fig. 1(b) for $\varepsilon_w=0.1$. Note also that $m_e$, the leveling 
off value of $m$, scales with $\varepsilon_a$,
and, as argued below, it scales with $\sigma^{-2}$ as well, in 
agreement with Fig. 1(c). Results for different cubic lattices are
shown in Fig. 1(c). The $\log m$ versus
$\log t$ slopes in the intermediate time regime are clearly lattice 
structure dependent. Much of Eqs. (2-5) is inferred from these
graphs.

For most of the rest of the paper, we try to understand these 
results. Let's first examine the physics of the waiting stage. We
assume the system is first cooled to some temperature $T_a$ that is 
above the ordering temperature $T_0$, but not infinite. We also
assume that
$k_BT_a\gtrsim U/10$. It then follows from Arrehnius' law,
$\tau =\tau_0 \exp (U/k_BT)$, that over barrier spin flipping readily 
takes place in the laboratory within a second's time if
$\tau_0\lesssim 10^{-5}$ s. Some correlation between spin $s_i$ and 
field $h_i$ at each site $i$ can therefore be established, but no
long-range order can develop if $T_a\gtrsim T_0$. Assume that either 
$T_a\gg T_0$ or that the time spent in
the waiting stage is so short that the probability density function 
(PDF) $p(h)$ that a randomly chosen site have field $h$ is
reasonably approximated by $p(h)\propto\exp (-h^2/2\sigma^2)$. On the 
other hand, the conditional PDF to find $\pm S$ given a field
$h$ acting on the spin fulfills, in equilibrium, $p(\pm S\mid 
h)\propto \exp (\pm  h/k_BT)$. Now, since the
joint probability density $p(\pm S,h)$ that, on a randomly chosen 
site, one find $h$ acting on $\pm S$ is in general given by
$p(\pm S,h)=p(\pm S\mid h)p(h)$,
\begin{equation}
p(\pm S,h)\propto e^{-(h\mp\sigma^2/k_BT)^2/2\sigma^2},
\label{x2}
\end{equation}
follows in equilibrium. Therefore, the mean energy is $- \sigma^2 
/2k_BT$. The replacement $
\sigma^2/2k_BT\rightarrow\varepsilon_a$, generalizes the above equation to
\begin{equation}
p(\pm S,h)\propto e^{-(h\mp 2\varepsilon_a)^2/2\sigma^2}
\end{equation}
for all times up to equilibration. Then, to first order in 
$\varepsilon_a/\sigma$,
\begin{equation}
p_\uparrow (h)-p_\downarrow (h)\simeq
\sqrt{2/\pi}\,h\varepsilon_a\sigma^{-3}e^{-h^2/2\sigma^2},
\label{pmp}
\end{equation}
where $p_\uparrow (h)=p(S,h)$ and $p_\downarrow (h)=p(-S,h)$.
All points for $[p_\uparrow (h)-p_\downarrow (h)]/(h\varepsilon_a)$ 
obtained from MC calculations collapse onto a single
curve in Fig. 2a, in agreement with Eq. (\ref{pmp}). Deviations, of 
higher order in $\varepsilon_a/\sigma$, from Eq.
(\ref{pmp}) do occur. They are within 10\% even if complete thermal 
equilibration is allowed to
take place as long as $T\gtrsim 10$, i.e., above approximately $4T_0$.
\begin{figure}
\includegraphics[width=8cm]{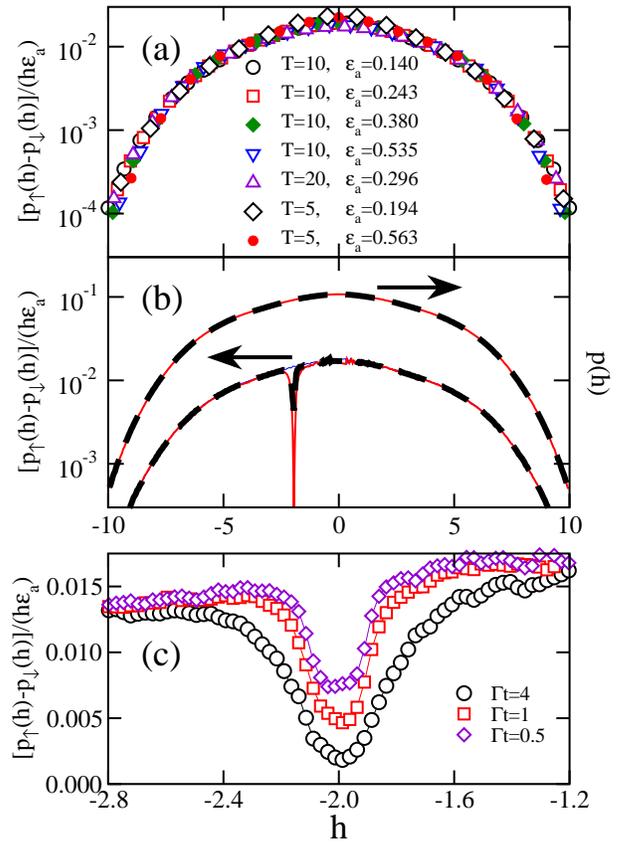}
\caption{$[p_\uparrow (h)-p_\downarrow (h)]/(h\varepsilon_a)$ for a 
system of $16\times 16\times 16$ spins at $T_a=10$
(recall $T_0\simeq 2.5$), before quenching. The shown temperatures 
are given in units of $\varepsilon_d/k_B$. The
system was cooled to these ``high'' temperatures from an infinite
temperature, and allowed to thermalize for a time, till the shown 
energies were reached. All points stand for averages over
$1.5\times 10^5$ runs. (b) $[p_\downarrow (h)-p_\uparrow 
(h)]/(h\varepsilon_a)$ and $p(h)\equiv
p_\downarrow (h)+p_\uparrow (h)$ at times $t=0$, $\Gamma t=0.62$ 
(discontinuous line), and $\Gamma t=2.19$ (continuous line), for the
same system as in (a), after the system was first thermalized at
$T=10$ till $\varepsilon_a\simeq 0.25$, was then further cooled to 
$T=1$, and $H=2$ was then applied.
A tunnel window $\varepsilon_w=0.1$ was enforced. All points in (b) 
stand for averages over $4\times 10^4$
histories. (c) Same as in (b) but for $\Gamma t=0.5$, $1$, and $4$; 
in addition,
while $t<0$, the system had been partially thermalized at $T=10$ till 
$\varepsilon_a=0.247$.}
\label{11}
\end{figure}

We now examine the time evolution of the system after abruptly 
cooling it, at time $t=0$, to a temperature below, roughly,
$ 0.2U/(Sk_B)$. Then, spin flips up to $\mid S_z\mid <\,S$ states can 
be neglected, and
tunneling through the ground state doublet is the only available path 
for spin reversals. Accordingly, spin flips are allowed only if
the spin's Zeeman energy is within the tunnel window. A field $H$ is 
applied for all $t>0$. Now, if either the system is in thermal
contact with a reservoir at a temperature such that 
$k_BT\gg\varepsilon_w$, or the energy is constant and sufficiently
high such that $k_BT_u\gg\varepsilon_w$, then
\begin{equation}
\dot{m}=2\Gamma \int dh\; \eta(H+h)f(h,t) ,
\label{uno}
\end{equation}
where $f(h,t)=p_\downarrow (h,t)-p_\uparrow (h,t)$.

It is worth pointing out that $-f(h,0)$ is given by Eq. (\ref{pmp}), 
and therefore $f(h,0)=-(2h\varepsilon_a/\sigma^2) p(h,0)$
at $t=0$, where $p(h,0)$ is the PDF regardless of spin orientation, 
i.e., $p(h)=p_\downarrow(h)+p_\uparrow (h)$.
However, $f(h,t)\propto hp(h,t)$ does {\it not} hold for $t>0$. This 
point is illustrated in Fig. 2(b), where MC results for
both $f(h,t)$ and $p(h,t)$ are shown for some non-zero times. 
$p(h,t)$ is invariant for times $\sim\Gamma$. As reported in
Ref.\cite{Alonso}, a sharp dip, of approximately
$\varepsilon_w$ half--width, does develop in $p(h)$, but only at much 
later times, and then not when
$k_BT\gg\varepsilon_w$. On the other hand, a hole does show up in 
$f(h,t)$ in Fig. 2(b), as in Wernsdorder's experiments
\cite{ww1}, performed at $T=40$ mK (which is roughly 10 times as 
large as $\varepsilon_w/k_B$) \cite{Alonso,ww1}.

The time development of the hole in $f(h,t)$ is illustrated in Fig. 
2(c). Note that the hole deepens, but its width remains
approximately constant while $\Gamma t\ll 1$, and then, under the 
conditions stated above Eq. (\ref{uno}), $\dot{f}=-2\Gamma f$.
Therefore, since $m$ equals the area coverd by the hole [i.e., 
$m(t)=\int dh\,[f(h,0)-f(h,t)]$],
\begin{equation}
m\simeq 2\varepsilon_wf(-H,0)(1-e^{-2\Gamma t})
\end{equation}
if $\Gamma t\ll 1$ and $\varepsilon_w\ll \sigma$. Using Eq. 
(\ref{pmp}), Eq. (3) follows. The value $m_e$ that $m(t)$ levels off 
to
after a sufficiently long time, that is, Eq. (5), can be estimated as 
follows. For
$\Gamma t\gtrsim 1$, $f(-H,t)\ll f(-H,0)$, and the hole only becomes 
broader, but it cannot become wider than the field
distribution $p(h)$. The final area covered by the hole is therefore 
$\sim 2f(-H,0)\sigma$, which is the estimated value of $m_e$,
in rough agreement with the expression for $m_e$ below Eq. (5).

More detailed considerations underlie Eq. (4). We have derived 
\cite{theory} the equation,
\begin{equation}
\dot{x}\simeq c_1\sqrt{\frac{2}{\pi}}-c_2\int_0^t 
d\tau\frac{2\varepsilon_w\dot{x}(\tau )}{\omega 
(t-\tau)+2\varepsilon_w},
\label{mfinal}
\end{equation}
where, $c_1$ and $c_2$ are constants to be specified below, $\omega 
(t-\tau )$ is the inverse of the PDF that the field $h$ on a
randomly chosen site be the same at times $t$ and $\tau$. Quantity 
$m$ follows from Eq. (\ref{mfinal}) letting
$m=x\varepsilon_a\varepsilon_wH/\sigma^3$, $c_1=4$, and $c_2=2$. 
$\omega (t-\tau )$ depends on $t-\tau $ through the
probability
$\phi(t-\tau )$ that a spin point in opposite directions at times $t$ 
and $\tau $. To make progress, we make the approximation
$\omega\simeq \min [2\pi h_0\phi, 2\sigma\sqrt{\pi}\sqrt{\phi} ]$. 
The approximation for
$\phi\ll 1$ follows from the Lorentzian PDF, of half width $h_0\phi$, 
that ensues when a small fraction $\phi$ of sites are
randomly occupied by spins \cite{PWA}. (The two factors of 2 come 
from the fact that flipping an already present spin $S$ is like
placing a
$2S$ spin on an unoccupied site.) The approximation for $\phi\sim 
1/2$ follows from the Gaussian distribution that holds then. In
between, the interpolation checks, within some
$10\%$, with our MC results. Finally, $\phi (t-\tau )$ must be 
evaluated. For this purpose, an
equation for the fraction of spins $n(t)$ that have flipped at least 
once before time $t$ is derived \cite{theory}.
It is Eq. (\ref{mfinal}) using $n=x\varepsilon_w/\sigma$, $c_1=1$ and 
$c_2=1$. We then use $\phi\approx n/2$. The functional
dependence of $F$ shown in Eq. (2) follows by careful inspection of 
these equations. Numerical calculations yield the curves shown in
Figs. 1(a), (b), and (c). The exponent $p$ depends on lattice structure,
through $\sigma /h_0$ in the expression for $\omega (\phi )$ above. 
The agreement with our MC results exhibited in Fig. 1(c)
is reassuring.

A couple of final remarks follow. Equations (2-5) are for magnetic 
systems that do not exchange energy with a heat bath (i.e.,
the lattice, usually) at very low temperatures. Then, $m(t)\rightarrow m_e$ as
$t\rightarrow \infty$. When heat exchange does takes place readily, 
as in some systems in Ref. \cite{metes}, then our simulations show
that $m(t)$ eventually crosses over from the value given by Eqs. (2-5) to
$m_{th}(t)\simeq 0.3(\varepsilon_w/\sigma)^3 H\Gamma t$, for 
$H\lesssim 1$. This happens when $m_{th}(t)$ becomes the larger of the
two. Later on, $m_{th}$ levels off to a quasi steady state value (not 
the final thermal equilibrium value), which depends on both
system shape and lattice structure, as expected. (If a field is not 
applied inmediately upon quenching, but later, the quasi steady
state value of
$m_{th}$ is also affected.) The $(\varepsilon_w/\sigma)^3$ dependence 
suggests that (see Ref. \cite{Ising}), in contrast with
constant energy magnetization processes, magnetic ordering takes 
place while the system magnetizes.

Summing up, we have given MC and theoretically based evidence to show 
that the $m\propto \sqrt{t}$ behavior
observed in experiments on Fe$_8$ clusters \cite{ww1} after quenching 
and applying a small field $H$ at $t=0$ is driven by
correlations which are previously established in the system while 
cooling to very low temperature. We have established that the
$\sqrt{t}$ behavior is not universal. More generally, $m(t)\propto t^p$, and
$p$ depends on lattice structure. The time range over which this 
behavior prevails, the value $m_e$ that the
magnetization later levels off to, and the crossover time to a final 
thermally--driven regime have been determined. More
specifically, Eq. (2-5) have been inferred from MC simulations and 
Eq. (\ref{mfinal}), and much of the relevant physics has been
explained.

We are grateful to F. Luis and W. Wernsdorfer for useful comments. 
The Ministerio de Ciencia y Tecnolog\'{\i}a of Spain
supported this work through grant No. BFM2000-0622.


\begin{thebibliography}{99}
\bibitem{Alonso2}In numerical work, we use $\sigma\simeq 3.83$, 
$4.01$, and $8.42$, which are the standard
deviation values (slightly larger than the corresponding rms values) for SC,
BCC, and FCC lattices, respectively.
\bibitem{theory}This can not be fitted here. It will be given elsewhere.
\end{thebibliography}
\end{document}